\documentclass{elsart}

\usepackage{epsfig}

\usepackage{amssymb}

\def\cf{{\it cf.~}}
\def\eg{{\it e.g.,~}}
\def\ie{{\it i.e.,~}}
\def\lsim{\raise0.3ex\hbox{$<$}\kern-0.75em{\lower0.65ex\hbox{$\sim$}}} 
\def\gsim{\raise0.3ex\hbox{$>$}\kern-0.75em{\lower0.65ex\hbox{$\sim$}}}

\def\cm3{~{\rm cm^{-3}}}

\def\euv{Extreme Ultra-violet}

\def\cf{{cf.~}}
\def\eg{{e.g.,~}}
\def\ie{{i.e.,~}}
\def\lsim{\raise0.3ex\hbox{$<$}\kern-0.75em{\lower0.65ex\hbox{$\sim$}}} 
\def\gsim{\raise0.3ex\hbox{$>$}\kern-0.75em{\lower0.65ex\hbox{$\sim$}}} 
\def\sc1{\raise2.1ex\hbox{\tiny $r\!\!=\!\!4$}\kern-0.95em{\hbox{$=$}}}

\def\cm3{~{\rm cm^{-3}}}

\def\hinv{$h^{-1}$}

\def\phat{\hat{p}}

\def\euv{extreme ultra-violet~}

\def\ltsima{$\; \buildrel < \over \sim \;$}
\def\simlt{\lower.5ex\hbox{\ltsima}}
\def\gtsima{$\; \buildrel > \over \sim \;$}
\def\simgt{\lower.5ex\hbox{\gtsima}}

\def\sc{{\rm Science\ }}

\def\lesssim{\mathrel{\hbox{\rlap{\hbox{\lower4pt\hbox{$\sim$}}}\hbox{$<$}}}}
\def\gtrsim{\mathrel{\hbox{\rlap{\hbox{\lower4pt\hbox{$\sim$}}}\hbox{$>$}}}}

\def\ion#1#2{#1$\;${\small\rm\@Roman{#2}}\relax}

\newcount\lecurrentfam
\def\LaTeX{\lecurrentfam=\the\fam \leavevmode L\raise.42ex
\hbox{$\fam\lecurrentfam\scriptstyle\kern-.3em A$}\kern-.15em\TeX}
\def\sizrpt{
(\fontname\the\font): em=\the\fontdimen6\font, ex=\the\fontdimen5\font
\typeout{
(\fontname\the\font): em=\the\fontdimen6\font, ex=\the\fontdimen5\font
}}
\def\plotfiddle#1#2#3#4#5#6#7{\centering \leavevmode
\vbox to#2{\rule{0pt}{#2}}
\includegraphics{#1}}
\def\refindent{\advance\leftskip by 24pt \parindent=-24pt}
\def\appgdef#1#2{%
 \@ifxundefined#1{\toks@{}}{\toks@\expandafter{#1}}%
 \toks@ii{#2}%
 \xdef#1{\the\toks@\the\toks@ii}%
}%
\def\@tablenotetext#1#2{%
 \vspace{.5ex}%
 {\noindent\llap{$^{#1}$}#2\par}%
}%
\def\@tablenotes#1{%
 \par
 \vspace{4.5ex}\footnoterule\vspace{.5ex}%
 {\footnotesize#1}%
}%
\def\@tablenotes@pptt#1{%
 \par
 \vspace{3.2ex}\footnoterule\vspace{.4ex}%
 {\footnotesize#1}%
}%
\AtBeginDocument{%
}%
\newcommand\tablenotetext[2]{%
 \appgdef\tblnote@list{\@tablenotetext{#1}{#2}}%
}%
\def\spew@tblnotes{%
 \@ifx@empty\tblnote@list{}{%
  \@tablenotes{\tblnote@list}%
  \global\let\tblnote@list\@empty
 }%
}%

\begin{document}

\begin{frontmatter}



\title{COSMOCR: 
A Numerical Code for Cosmic Ray Studies in Computational Cosmology}


\author{Francesco Miniati}

\address{Max-Planck-Institut f{\"u}r Astro\-phy\-sik,
Karl-Schwarz\-schild\--Str. 1, D-85741 Gar\-ching, Germany}
\address{School of Physics and Astronomy, University of Minnesota,
    Minneapolis, MN 55455}

\begin{abstract}

We present COSMOCR, a numerical code for the investigation of
cosmic ray related studies in computational cosmology.
The code follows the diffusive
shock acceleration, the mechanical and radiative energy
losses and the spatial transport of the supra-thermal particles
in cosmic environment.
Primary cosmic ray electrons and ions are injected at
shocks according to the thermal leakage prescription.
Secondary electrons are continuously injected as a results
of p-p inelastic collisions of primary cosmic ray ions and 
thermal background nuclei. 
The code consists of a conservative, finite volume method
with a power-law sub-grid model in momentum space.
Two slightly different schemes are implemented 
depending on the stiffness of the cooling terms.
Comparisons of numerical results with analytical solution 
for a number of tests of direct interest show remarkable 
performance of the present code.

\end{abstract}


\end{frontmatter}


\section{Introduction} \label{intro.se}

Structure formation is a major research area in cosmology and
a powerful observational constraint 
to discriminate among the viable cosmological models
\cite{bops99}.
For example, the evolution of the larger structures
such as Galaxy Clusters, 
being very sensitive to the underlying mass content of the Universe,
is used to evaluate the density parameter $\Omega_m$ for matter 
\cite{wnef93,ostrk93}.
Numerical simulations have proven an invaluable tool and 
have qualitatively confirmed our theoretical
framework for the mechanism of structure formation \cite{peebles93}.
Today, in the era of high precision cosmology, a substantial part 
of the efforts are directed toward the development of 
a coherent and quantitative picture. 
This must be inclusive of the feedback 
from forming structure (\eg stars, galaxies, and active galactic nuclei),
which strongly affects the evolution of the observable universe.

In this context, cosmic-ray (CR hereafter) pressure has been recently recognized 
as a possible source of significant dynamical feedback \cite{mint00}.
Indeed, during the hierarchical process of structure formation,
supersonic gas
in-fall and merging events invariably generate powerful, 
large and long-lived shock waves \cite{suze72s,bert85b,ryka97a,minetal00}.
These should produce copious amounts of CRs, by way of diffusive shock
acceleration \cite{bbdgpm91},
including both electrons and ions.
In addition, the post-shock gas and diffusively trapped CRs are
mostly advected into
non-expanding regions, such as filaments and clusters. 
It turns out that the energy of most of the CR-protons is only marginally
affected by radiative losses during a Hubble time. 
It is possible, then, that
the latter might accumulate inside large
forming structures, storing up a substantial fraction 
of the total pressure there \cite{bbp97}. In addition to cosmic shocks
other sources of CRs are also possible. 
These include active galactic nuclei, supernovae and stellar winds, 
all of which are
candidate for important contributions to the 
total population of CRs in cosmic structures \cite{jmrk00}.
There is growing observational evidence that significant 
non-thermal activity takes place in clusters of
galaxies. This evidence 
is provided by extended sources of radio
emission, namely radio halos and relics.
From its spectral properties and, sometimes,
polarization signatures the radio
emission is interpreted as synchrotron radiation,
implying the presence of 
relativistic cosmic-ray electrons and magnetic fields
\cite{kkgv89,kkdl90,giova93,deissetal97,gife00}.
Also, there have been claims of 
detection of radiation in excess to what is
expected from the hot, thermal X-ray emitting intra 
cluster medium, both in the hard X-ray
band above $\sim 10$ KeV \cite{fufeetal99,kaastraetal99} and
perhaps even in the
extreme ultra-violet
\cite{lieuetal96a,lieuetal96b,mll98,libomi99,bobeko99,bolimi00,beboko00}.
The importance of this non-thermal component and its 
cosmological implications will be discussed elsewhere
\cite{mjkr01,mrkj01}.

In this paper we describe 
a numerical code that allows a treatment 
of the evolution of various CR populations (\ie protons as well as
primary and secondary electrons) in computational cosmology. 
This code is instrumental for a detailed investigation 
of the aforementioned issues connected with the non-thermal 
activity associated with the formation of the large scale 
structure. 
In \S \ref{nuch.se} we outline the difficulty of the 
numerical treatment of CRs in this context and 
a strategy for useful applications. 
The code is extensively described in \S \ref{crco.se}. 
There we present the algorithm for the treatment 
of CR ions and electrons including their advection 
in both momentum (\S\S  \ref{admo.se}, \ref{admoe.se}) and physical space (\S \ref{adph.se}). 
In addition we describe the inclusion of 
two source mechanisms, namely injection at shocks (\S \ref{inj.se}) 
and production of secondary electrons (\S \ref{prosec.se}).
Finally, in \S \ref{tere.se} we present some numerical experiments
to test the code performance. 

\section{Challenges and Approach} \label{nuch.se}

The evolution of the 
supra-thermal particles, \ie CR 
electrons and ions is described by the diffusion-convection 
equation \cite{skill75a}. In 
comoving coordinates the latter takes the form
\begin{eqnarray}  \nonumber
\frac{\partial f}{\partial t} & = & 
- \frac{1}{a}\;{\bf u\cdot \nabla} f + 
\frac{1}{a^2}\;{\bf\nabla (\kappa\nabla }f) \\&  & \nonumber
+\frac{1}{a}
\left(\dot{a}+ \frac{1}{3}\, {\bf \nabla \cdot u}\right)
\: p\,\frac{\partial f}{\partial p} + \frac{1}{p^2} \, 
\frac{\partial}{\partial p} \left( p^2 \left[b_\ell(p)\,f + 
D_p \frac{\partial f}{\partial p} \right] \right)+ \\&  &
+ j({\bf x},p). \label{dce1.eq}
\end{eqnarray}
\noindent
where the gradient is with respect to comoving coordinates, ${\bf x}$,
$a(t)$ is the expansion parameter of the universe (such that ${\bf r} = a {\bf x}$ 
is the physical length), ${\bf u}= a(t) \dot{\bf x}$ is the peculiar velocity (\ie not inclusive
of the Hubble expansion) and $f({\bf x},p)$ is the isotropic part of the 
particle distribution function in comoving units.
Here $\kappa(p)$ and $D_p(p)$ are the diffusion coefficients in comoving
coordinates and momentum space, respectively. 
We recall that on the right-hand side of the 
above equation, the first line contains the spatial terms
of advection and diffusion respectively; the second line
includes the adiabatic losses due to cosmic ($\propto \dot{a}$)
and peculiar ($\propto {\bf \nabla \cdot u}$) expansion, other
mechanical and radiative losses 
[$\propto b_\ell(p)$; see eq. (\ref{radlos.se})]
and the second-order Fermi mechanism
($\propto D_p$); and the third line [$j({\bf x},p)$] represent the comoving source term,
which accounts for either fresh CRs 
injection at shock or secondary production.
Eq. (\ref{dce1.eq}) holds only for sub-horizon scales and, more importantly,
in the non-relativistic limit. 
These approximation are completely satisfactory for the investigations
of the problems outlined in the introduction. On the other hand, shock 
acceleration in relativistic flows such as those occurring in relativistic jets,
gamma-ray bursts and the like, needs to be treated differently
\cite{peacock81,hedr88,kgga00}.

The acceleration and transport of CR 
electrons and ions, described by the above equation,
involve physical scales that 
must be carefully considered 
when a numerical approach is undertaken.
The major difficulty arises in the attempt to model
the behavior of supra-thermal particles nearby shocks. 
There,
the smallest length scale of interest is that of the shock thickness,
$\ell_s$, which, for
a collision-less shock, is typically of order 
a few times the Larmor radius of a thermal proton 
\cite{stotsu85,tsusto85}, \ie
\begin{equation}
\ell_s \sim
r_L \sim  \frac{pc}{ZeB} \sim \;\left(\frac{T}{10^8\mbox{K}}\right)
\left(\frac{B}{0.1\mu\mbox{G}}\right)^{-1} 10^{11} ~~~ \mbox{cm}.
\end{equation}
Here $T$ is the post-shock temperature and 
$B$ is the magnetic field strength.
Particles injected at shocks
are thought to be pulled out from the high energy tail of the
thermal pool and have energy and a Larmor radius 
a few times above the thermal values
\cite{elmopa90,kajo91,joka93,beyeks94,beyeks96}. 
At even higher energy, CRs have 
a Larmor radius much larger than the shock thickness
and, therefore, unlike the thermal particles, are unaffected 
by the local shock transition.
In the vicinity of a shock the dynamics of
these supra-thermal particles is 
dominated by the effects of advection and diffusion,
which determine the energy gain
and escape probability for the particles \cite{drury83}. 
In particular, the spatial distribution is characterized by a 
diffusive scale-length $\lambda_d(p)= \kappa(p)/u_s$ (where $u_s$ is
the shock velocity), which determines the 
distance upstream of the shock to which a particle can propagate 
by diffusing against the advective flow.
For a correct numerical solution of eq. (\ref{dce1.eq}),
the hierarchy of the aforementioned scale lengths for all relevant
momentum,
namely that $\ell_s <  \lambda_d(p) \lsim \lambda_d(p_{max})$, must be properly
reproduced in the numerical grid \cite{jre99}. 
This demands a full resolution of the diffusion length,
typically $\Delta x \lsim \lambda_d/10$, as the shock discontinuity is
usually spread over 2-3 grid zones in a numerical simulation. 
When this spatial 
resolution is achieved, the physical roles of diffusion and
advection in eq. (\ref{dce1.eq}) are well accounted for, and
the property of continuity 
of the complete distribution function,
at the shock surface can be reproduced \cite{drury83}.
Notice that the continuity of the distribution function
across the shock (in the shock frame),
justified by the fact that the supra-thermal particles
do not experience the discontinuity,
is the very condition that determines the power
law behavior of the solution $f({\bf x},p)$ emerging from a shock wave 
\cite{drury83}.
Otherwise, the finite, numerical size of the shock, $\Delta x_s$, 
induces an error in the solution of eq. 
(\ref{dce1.eq}) of order $\Delta x_s/\lambda_d$ \cite{jre99}.

However, resolving spatial scales down to the 
sub-diffusion length of the supra-thermal particles turns 
out to be an extremely demanding task. In fact, 
if we assume for simplicity 
the case of Bohm diffusion, $\kappa = \kappa_B = \frac{1}{3} r_L v$, 
with $r_L$ and $v\sim c$, the particle Larmor radius and 
velocity, respectively, we find
\begin{equation}
\lambda_d(p) = \frac{\kappa_B(p)}{u_s} =  1.1\; \left(\frac{E}{\mbox{GeV}}\right)
\left(\frac{B}{0.1\mu\mbox{G}}\right)^{-1}\;
\left(\frac{u_s}{10^2\mbox{Km\,s}^{-1}}\right)^{-1} 10^{-2} ~~~\mbox{pc}.
\end{equation}
where $ E\simeq$ pc is the particle energy (in the relativistic limit)
and $u_s$ is the shock speed. This means that for electrons
with energy $\lesssim  1-10$ GeV, 
which are of interest for the non-thermal emission in 
galaxy clusters, the diffusion
length is $\lesssim 10^{-2}$pc and should be even smaller at injection
energies. Considered the
importance of including the cosmic structure on a scale $\gsim$
50\hinv Mpc \cite{osce96}, the spatial scales to be resolved extend
over more than 10 orders of magnitude!
Therefore, we can typically only afford grid sizes 
$ \Delta x \gg \lambda_d (p)$ and 
following the full dynamics of the CRs injection 
and acceleration 
at cosmological shocks becomes impractical.
However, although limited by the above restrictions, 
one can still attempt to find a useful numerical 
treatment of the CR particles. In particular, 
the instance that 
$\lambda_d (p)/ \Delta x \ll 1$ also implies that
the diffusion time at shocks, $\tau_d(p)= \lambda_d / u_s \le 
(\lambda_d (p)/ \Delta x)\;\Delta t \ll \Delta t$, \ie much 
smaller than the 
computational time-step \cite{jre99}.
The diffusive shock acceleration time, $\tau_{acc}$,
for a strong shock ($r \equiv u_1/u_2 \simeq 4$) is
\begin{eqnarray} 
\nonumber
\tau_{acc}(p) & = &
\frac{3}{u_1 -u_2}\,\left(\frac{\kappa_1}{u_1}+\frac{\kappa_2}{u_2}\right)=
3r\;\left(\frac{r+1}{r-1}\right)\;\frac{\kappa(p)}{u_1^2} \; \simeq \;
20 \;\tau_d(p)
\\&  =  &
 21.1\; \left(\frac{E}{\mbox{GeV}}\right)
\left(\frac{B}{0.1\mu\mbox{G}}\right)^{-1}\;
\left(\frac{u_s}{10^2\mbox{Km\,s}^{-1}}\right)^{-1}\; \mbox{yr}
\end{eqnarray}
where $r$ is the shock compression
ratio, subscripts 1 and 2 indicate 
values upstream and downstream of the shock, respectively, and
for simplicity we have assumed $\kappa_1=\kappa_2$.
Since the time-step for a 
cosmological simulation is of order of $10^8$ yr, 
with the above representative values of magnetic field strength and
shock speed,
$\Delta t \gg \tau_{acc} $ up to a CR energy $E\simeq 10^{15}$eV. 
This energy is well beyond the range of energies of interest 
for CR electrons in clusters of galaxies and should 
include most of the CR ions dynamically relevant in those
environments. In fact at energies much larger than that, both
energy losses \cite{masc94} 
and diffusive escape \cite{bbp97} become important 
reducing the population of CRs with $E\gsim 10^{15}$eV.
Thus, for both CR electrons and ions within the energy
range derived above, we will assume 
that shock acceleration is {\it instantaneous} \cite{kajo91,drury91} and
that a downstream steady-state solution is generated
in accord with the properties of the shock
\cite{jre99}. 
The simplest way to implement this prescription
({\it acceleration}) is to ``inject'' in correspondence of shock fronts
a test particle distribution of CRs (\cf \cite{drury83})
%
%
\begin{equation} \label{finj.eq}
f(p)  =   f_{inj} \; \left(\frac{p}{p_{inj}}\right)^{-q_s}
\end{equation}
%
where the slope is given by 
\begin{equation}  \label{slop.eq}
q_s  =   \frac{3r}{r-1}.
\end{equation}
Here 
the injection momentum $p_{inj}$ is determined by the post-shock
temperature (see in \S \ref{inj.se} for details.)
In addition, when a pre-existing population of CRs, $f_0(p)$, 
encounters a shock, the particle distribution is modified 
({\it re-acceleration}) only if 
the log-slope of the current distribution,
$ q_0  =  -\partial\ln  f_0/\partial \ln p$,
%
%
%
is larger than that determined by the shock compression ratio
[eq. (\ref{slop.eq})] (\cf \cite{drury83}).
In general,
the injected and re-accelerated particle distributions
could also be affected by the nonlinear 
feed-back of the CRs themselves \cite{beel99}. 
In this case the distribution function deviates
from the simple power-law given in eq.(\ref{finj.eq}).

Once the supra-thermal particles have been produced
at shocks, they are carried through ``smooth'' flows
together with the background gas (before another encounter with a shock).
The circumstance that $\lambda_d\ll \Delta x$ suggests that over 
the scale $\Delta x$ of the numerical grid, the propagation of the 
particles is dominated by the advection term. In fact, the relative 
contribution of advection and diffusion in eq. (\ref{dce1.eq})
is of order $k/u_s\,\Delta x=\lambda_d/ \Delta x\ll 1$. Thus,
for the spatial transport of the CR particles in 
shock-less regions, the diffusion term is negligible and 
can be dismissed in eq. (\ref{dce1.eq})
\cite{jre99}. 

\section{The Cosmic Rays Code} \label{crco.se}

\subsection{Advection in Momentum Space} \label{admo.se}

The difficulty of a numerical treatment of eq. (\ref{dce1.eq})
is aggravated by the fact in practice
we are allowed to use only a ``handful'' of grid zones 
in momentum space. To illustrate, in the hypothesis that the number
of grid zones used for the one-dimension momentum variable 
is the same as for the spatial variables,
a $N^4$ single precision
array for a CR population would
require already about 17Gb for $N=256$.
The number of operations to be computed at each 
cycle grows consequently ($\propto N^4$) which is paid 
at the high cost of a low speed code. Finally, we must not
forget that the data-set output
by the simulation becomes huge, rendering the 
post-computation analysis memory intensive and difficult to
handle even on well equipped workstations.  

With only a few grid zones spanning the momentum space, 
a sub-grid model of the distribution function is necessary.
For this purpose, 
in the following we adopt the general approach first
developed in ref. \cite{jre99}, although some aspects of
the implementation of the numerical schemes below are different 
from those authors. In particular we compute the flux in momentum
space exactly up to a logarithmic term
and provide an alternative scheme for the stiff losses case.
We sub-divide the region of
$p$-space of interest into $N_p$ logarithmically spaced 
momentum {\it bins},
bounded by $p_0,\ldots,p_{N_p}$. The width of the bins is
expressed on a log scale as $\Delta \log p \equiv \Delta w_i
= \log(p_{i}/p_{i-1})$, which is convenient (although not necessary) 
to take as constant.
On this grid, we approximate $f(p)$ with the following
piece-wise power law:
\begin{equation} \label{piewi.eq}
f(p) = f_i \; \left(\frac{p}{p_{i-1}}\right)^{-q_i}
\end{equation}
where $f_i$ and $q_i$ are the normalization and logarithmic slope 
for the $\imath_{th}$ momentum bin. 
This approximation is valid when 
$q(p)$ changes slowly inside each bin, which holds true
as long as energy losses do not produce strong curvature
in the distribution function $f(p)$.
Then, integration over the $\imath_{th}$ momentum bin of
eq. (\ref{dce1.eq}) multiplied by a factor $4\pi\,p^2$,
gives
\begin{eqnarray} \nonumber
\frac{\partial n_i}{\partial t}& = &- \frac{1}{a}\;{\bf \nabla\cdot( u}\,n_i) 
+ \frac{1}{a^2}\;{\bf\nabla (\langle \kappa \rangle \nabla }n_i) + \\& & + 
\frac{1}{a} \left\{ \left[ 
\left(\dot{a}+ \frac{1}{3}\, {\bf \nabla \cdot u}\right)\: p\,
+ a\,\left(b_\ell(p)+ D_p \frac{\partial \log f}{\partial p}\right)
\right] \;4\pi p^2 f(p)\right\}_{p_{i-1}}^{p_i}  +  \nonumber  \\& & + 
Q_i \label{dce2.eq}
\end{eqnarray}
where
\begin{eqnarray} \label{nbin.eq}
n_i & = & \int_{p_{i-1}}^{p_i} 4\pi\,p^2 \;f(p)\; dp = 4\pi \;f_i\;p_{i-1}^3\;
\frac{\left(\frac{p_i}{p_{i-1}}\right)^{3-q_i}-1}{3-q_i} \\ \label{kave.eq}
\langle \kappa \rangle _i & = & 
\frac{
\int_{p_{i-1}}^{p_i} p^2 \kappa \nabla f\; dp 
}{
\int_{p_{i-1}}^{p_i} p^2 \nabla f\; dp
}
 \\ 
Q_i & = & \int_{p_{i-1}}^{p_i} 4\pi\;p^2 \;j(p)\; dp  \label{source.eq};
\end{eqnarray}
and where we have used eq. (\ref{piewi.eq}) to derive the expression 
(\ref{nbin.eq}).
As already pointed out, away from shocks we can neglect the diffusive term.
Furthermore, we henceforth drop the second order Fermi term and focus on
the solution of the equation
\begin{equation} 
\frac{\partial n_i}{\partial \tau} = - {\bf \nabla\cdot( u}\,n_i)  + 
\left[ 
b(p) \;4\pi \;p^2\; f(p)\right] _{p_{i-1}}^{p_i}  +Q_i \label{dce3.eq}
\end{equation}
where $\tau=t/a$ and we have introduced
\begin{equation}  \label{bptdef.eq}
b(p) \equiv -\left(\frac{dp}{d\tau}\right)_{tot}
= \left(\dot{a}+ \frac{1}{3}\, {\bf \nabla \cdot u}\right)\: p\,
+ a\,b_\ell(p)
\end{equation}
which includes, in addition to other energy loss terms
($b_\ell(p)$; see \S \ref{radlos.se}), the adiabatic terms due to 
cosmic expansion and fluid divergence (convergence). 
For the sake of clarity,
we shall address now separately the details regarding the 
integration of each term of eq. (\ref{dce3.eq}). To begin with, we
will focus on the contribution due to advection in momentum space
responsible for the change
\begin{equation} 
\frac{\partial n_i}{\partial \tau} =\left[ 
b(p) \;4\pi \;p^2\; f(p)\right] _{p_{i-1}}^{p_i} \label{dcepiece1.eq}
\end{equation}
Time integration of eq. (\ref{dcepiece1.eq}) yields
\begin{equation} \label{dnpadv.eq}
n^{\tau+\Delta \tau}_i - n^\tau_i = - \Delta \tau
\left(\Phi_{i}^p-\Phi_{i-1}^p\right) 
\end{equation}
where
\begin{equation}  \label{fluxp.eq}
\Phi_i^p  = - \frac{1}{\Delta \tau}
\int_\tau^{\tau+\Delta \tau} b(p)\;4\pi \;p^2\; 
f(\tau^\prime,p)\;\vert_{p_i}\; d\tau^\prime,
\end{equation}
is the time-averaged flux
and the integrand is evaluated at the cell boundary $p_i$.
Recalling eq. (\ref{piewi.eq}),
the above integration is readily carried out after using eq.
(\ref{bptdef.eq}) to rewrite it as
\begin{equation}  \label{fluxp1.eq}
\Phi_i^p  = \frac{4\pi}{\Delta \tau}
\int_{p_i}^{p_u} \;p^2\; f_j(p)\; dp
\end{equation}
where 
\begin{equation} \label{jdef.eq}
j = 
   \left\{ \begin{array}{lll}  i+1 & \mbox{if} &  b(p_i)>0\\ 
   i &  \mbox{if} &  b(p_i)\le
   0 \end{array} \right.
\end{equation}
and $p_u$ is the upstream momentum, solution of the integral 
equation
\begin{equation}   \label{pu.eq}
\Delta \tau = 
= - \int^{p_u}_{p_i}  \frac{dp }{b(p)}
\end{equation}

After updating $n_i$ based on (\ref{dnpadv.eq}), and including
the contributions from spatial advection and injections 
(detailed below), the sub-grid model of the distribution 
function (\ref{piewi.eq}) is reconstructed by
solving for $f_i$ and $q_i$ at each computational cell. 
For each new value $n_i$, $f_i$ and $q_i$ are related by
eq. (\ref{nbin.eq}), so one additional constraints is necessary.
For the second relation 
we assume that the curvature of the spectrum is constant,
\ie $q_{i+1} -q_i =q_i -q_{i-1}$ (\cf \cite{jre99} for 
more detail on this). This is a sensible
approximation when cooling is weak. It is, therefore,
very suitable for cosmic
ray ions, which, in a cosmological context and for the energy range of 
interest here, suffer only minor losses even during a Hubble time.
However, for the case of electrons, which cool much more rapidly, 
this approximation breaks down and additional measures must be taken,
which we address below.

\subsection{Alternative Scheme for Fast Cooling Electrons} \label{admoe.se}

The assumption of a smooth CR energy distribution 
(constant spectrum curvature),
which allows one to infer the slope 
of the distribution function from the particle 
number density in each bin, 
breaks down in the presence of strong cooling. 
This is an important issue for electrons,
since the cooling times 
associated with inverse Compton and synchrotron losses 
are much shorter than the time-scale of a cosmological 
simulation.
It is important to notice that the effect of strong
inverse Compton or synchrotron cooling is that of producing 
a cut-off in the particle distribution, not just a mere 
steepening of it. Since each momentum bin spans 
a considerable range of values, it is often the case that
the cut-off falls in the middle of a bin, so that 
only the part of the bin below the energy cut-off is populated, while
upper part is 
completely depleted. This situation is likely to occur for
any of the momentum bins, not just those at the highest energies,
because, due to additional Coulomb losses 
(see \S \ref{radlos.se}), the cooling times for the
electrons at all energies of interest 
are always smaller than or comparable to the Hubble time. 
For the same reason, except around shocks where 
they are continually injected, primary CR electrons
in most of the simulated volume only occupy a very 
few of the lower energy momentum bins. 
This adds a non-trivial complication because at least 
three momentum bins are necessary for the reconstruction
scheme mentioned in the previous section to work \cite{jre99}.

In order to circumvent these difficulties,
we have devised an alternative scheme to be employed 
for the electrons only. 
The idea consists of replacing the ``constant curvature'' assumption 
with a different condition that allows us to reconstruct both 
normalization, $f_i$, and slope, $q_i$, of the piecewise power-law 
distribution function (\ref{piewi.eq}) in each bin. 
A natural constraint is provided by the physical condition that the 
total energy be conserved, except for the explicit sources and sinks.
The corresponding equation is derived by
taking one {\it moment} of the transport equation (\ref{dce1.eq}).
Thus, after multiplying by a factor $4\pi\,p^2\,T(p)$, where 
$T(p)=(\gamma -1)\,m_e\,c^2$ is the particle kinetic energy
(where $\gamma = 1/\sqrt{1-(v/c)^2}$ is Lorentz factor),
and integrating over the $i_{th}$ momentum bin, we have
\begin{eqnarray} \nonumber
\frac{\partial \varepsilon_i}{\partial t}& = 
&- \frac{1}{a}\;{\bf \nabla\cdot( u}\,\varepsilon_i) 
+ \frac{1}{a^2}\;{\bf\nabla (\langle \kappa_T\rangle_i \nabla }\varepsilon_i) \\& & + 
\frac{1}{a} \left\{ \left[ 
\left(\dot{a}+ \frac{1}{3}\, {\bf \nabla \cdot u}\right)\: p\,
+ a\,b_\ell(p)
\right] \;4\pi p^2 f(p)\,T(p)\right\}_{p_{i-1}}^{p_i}  \nonumber  \\& & 
- \frac{1}{a} \,\int_{p_{i-1}}^{p_i} \left[ \left(\dot{a}+ \frac{1}{3}\, 
{\bf \nabla \cdot u}\right)\: p\, + a\,b_\ell(p)
\right] \;4\pi p^2 f(p)\,\frac{p}{\sqrt{m_e^2\,c^2+p^2}}\;dp
\nonumber \\& & + 
S_i \label{edce.eq}.
\end{eqnarray}
As before, we have dropped the second order Fermi term, 
and introduced the following definitions
\begin{eqnarray} \label{ebin.eq}
\varepsilon_i & = & 
\int_{p_{i-1}}^{p_i} 4\pi c\,p^2 \;f(p)\,T(p)\; dp 
= 4\pi c \;f_i\;p_{i-1}^4\;
\frac{\left(\frac{p_i}{p_{i-1}}\right)^{4-q_i}-1}{4-q_i} \\ 
\label{ekave.eq}
\langle \kappa_T \rangle _i & = & 
\frac{
\int_{p_{i-1}}^{p_i} \kappa\,  p^2\,(\nabla f)\,T(p)\; dp 
}{\int_{p_{i-1}}^{p_i} p^2 \,(\nabla f)\,T(p)\; dp}
\\ 
S_i & = & \int_{p_{i-1}}^{p_i} 4\pi\;p^2 \;j(p)\,T(p)\; dp  \label{esource.eq},
\end{eqnarray}
where eq. (\ref{piewi.eq}) has been used and 
the sub-relativistic contribution has been ignored
in order to derive the last expression in eq. (\ref{ebin.eq}).
The first two terms on the right hand side of eq. (\ref{edce.eq})
represent the usual spatial advection and diffusion respectively. 
The third and fifth terms account for 
advection in momentum space and external energy 
source respectively, 
analogous to the corresponding third and fourth terms in 
eq. (\ref{dce2.eq}). Finally, in the fourth term on the right hand side 
we combine, for reasons of numerical convenience, 
both the CR pressure and the sink contributions.

As before, the diffusion term can be neglected, whereas both
spatial advection and energy injection will be treated in the next sections.
The changes affecting the distribution in momentum space are 
then provided by the third and fourth terms according to
\begin{equation} \label{edcepiece1.eq}
\frac{\partial \varepsilon_i}{\partial \tau} =\left[ 
b(p) \;4\pi \;p^2\; f(p)\,T(p) \right] _{p_{i-1}}^{p_i} -
\int_{p_{i-1}}^{p_i} b(p)
\;\frac{4\pi p^3 f(p)}{\sqrt{m_e^2\,c^2+p^2}}\;dp,
\end{equation}
where $\tau=t/a$ and $b(p)$ is defined by (\ref{bptdef.eq}).
For the range of energy of interest
here $p\gg m_e\,c$. Then, by using 
the prescription in (\ref{piewi.eq}) for $f(p)$,
the second term in eq. (\ref{edcepiece1.eq})
can be rewritten as $\varepsilon_i R_i(q_i,p_{i-1})$, where:
\begin{equation} \label{rterm.eq}
R_i(q_i,p_{i-1}) = 
\; \int_{p_{i-1}}^{p_i} b(p)\;\frac{p^{3-q_i}}{\sqrt{m_e^2\,c^2+p^2}}\;dp
~~/ ~~\int_{p_{i-1}}^{p_i} \, p^{2-q_i} \, T(p)\; dp
.
\end{equation}
Thus, after integration over a time-step
eq. (\ref{edcepiece1.eq}) reads
\begin{equation} \label{ednpadv.eq}
\varepsilon^{\tau+\Delta \tau}_i 
\;\left(1+\frac{\Delta \tau}{2}R_i\right) = 
 \varepsilon^\tau_i \;\left(1-\frac{\Delta \tau}{2}R_i\right) - \Delta \tau
\left(\Phi_{i}^\varepsilon-\Phi_{i-1}^\varepsilon\right)
\end{equation}
where the term $\propto R_i$ has been integrated implicitly and
\begin{equation}  \label{efluxp.eq}
\Phi_i^\varepsilon  = - \frac{1}{\Delta \tau}
\int_\tau^{\tau+\Delta \tau} b(p)\;4\pi \;p^2\; 
f_i(\tau^\prime,p)\,T(p)\;\vert_{p_i}\; d\tau^\prime,
\end{equation}
is the time-averaged flux whose 
integrand part is evaluated at the cell boundary $p_i$.
In analogy with the previous section,
by using the definition (\ref{piewi.eq}) and eq.
(\ref{bptdef.eq})
the above integral is readily rewritten as
\begin{equation}  \label{efluxp1.eq}
\Phi_i^\varepsilon  = \frac{4\pi}{\Delta \tau}
\int_{p_i}^{p_u} \;p^2\; f_j(p)\,T(p)\; dp
\end{equation}
where the index $j$ and the upper integration limit, $p_u$,
are defined in (\ref{jdef.eq}) and (\ref{pu.eq}) respectively.
In addition, the maximum momentum of the electrons distribution,
$p_{cut}$, 
is followed explicitly by means of equation (\ref{bptdef.eq}),
which can be easily integrated analytically \cite{karda62}
for the cooling mechanisms of interest here.
Therefore, in this scheme in eq. (\ref{efluxp1.eq}) 
and eq. (\ref{fluxp1.eq}) we adopt
the minimum between $p_u$ and $p_{cut}$.
Following explicitly the value of $p_{cut}$, \ie
the high energy cut-off of the distribution, allows us
to describe the electrons in the last momentum bin by 
a normalization and slope appropriate for the still 
populated portion of that bin.

After accounting for the advection in physical space and
source terms treated in the following sections,
we have information on both the particle number density and
kinetic energy at each momentum bin. Taking the ratio 
of energy to number density for each bin, we find for $p_i \gg mc$. 
\begin{equation}
\frac{\varepsilon_i}{n_ip_{i-1} c} =
\frac{3-q_i}{\left(\frac{p_i}{p_{i-1}}\right)^{3-q_i}-1} \,
\frac{\left(\frac{p_i}{p_{i-1}}\right)^{4-q_i}-1}{4-q_i}
\end{equation}
which can be solved in the unknown $q_i$
through Newton-like method with only a few iterations.
The value of $f_i$, which is actually not explicitly needed in any integration 
step, can easily be computed from either definition
(\ref{nbin.eq}) or (\ref{ebin.eq}). 
In this method both $f_i$ and $q_i$ are derived from quantities
pertaining exclusively to the $i_{th}$ bin and 
no information about the adjacent bins is required. 
Thus the scheme works with even one single cell or 
a fraction of it (since we keep track of $p_{cut}$ which can 
even be smaller than $p_1$).
This allows us to follow for cosmological times the low energy  
electrons with $\gamma \sim 300$. The electrons in this energy 
range have the longest lifetime against energy losses in a 
cosmological environment \cite{sarazin99}.
Thus, about $10^9$yr after a CR distribution has been produced
in a shock, these are the only electrons left. However, 
It is important to keep track of them, even if most of the electrons
at higher or lower energy have been depleted,  
because they can contribute to 
the \euv radiation detected in clusters of galaxies.

\subsection{Constraint on the Time-step} \label{cour.se}

In order for the scheme described in \S \ref{admo.se} and \S \ref{admoe.se} 
to behave properly the time 
step $\Delta \tau$ must be such that
$|\log\frac{p_u}{p_i} | \le  \epsilon_{CFL} \Delta w $
where $\epsilon_{CFL}\le 1$ is a Courant-like number. 
Given the very long cooling times
for the proton CRs compared to the typical 
cosmological time-step,
there is no need to enforce the above
condition for this case.
For the electrons, however, we find that in order to follow
accurately the evolution near the high energy cut-off a 
limit on the time-step
$\Delta \tau \lesssim 0.1 \tau_{sync+IC}$ should be adopted.

\subsection{Advection in Physical Space} \label{adph.se}

While we have focused so far almost exclusively on the integration
of the diffusion-convection equation in momentum space,
the full evolution of the CR particles must also include 
the effects of spatial transport. 
Since we neglect the spatial diffusion 
for reasons already pointed out in \S \ref{nuch.se}, 
the main contribution in this respect is provided by
the advection terms. In eq. 
(\ref{dce3.eq}) and (\ref{edce.eq}),
these are of the form
\begin{equation} 
\frac{\partial c_i}{\partial \tau} 
= - {\bf \nabla \cdot } ( {\bf u} c_i )
\label{dcepiece2.eq}
\end{equation}
with $c_i = n_i $ or $\varepsilon_i$ respectively.
These terms are integrated 
by means of a van Leer method \cite{vanleer73}, which is a 
conservative, second order accurate and 
Total Variation Diminishing scheme. 
The scheme allows integration of the 
advection terms in one dimension and the full integration
along all three different 
directions is achieved by
directional splitting \cite{strang68}. 
As a general feature 
of conservative, Godunov-like schemes \cite{godunov59}, 
the quantity that is being
updated is the volume-average over the computational cell
\footnote{We point out, for the sake of completeness, 
that because of this volume-averaging step, the distribution 
function $f$ employed in \S \ref{admoe.se} 
should be read as $\bar{f}$, \ie its volume average.}
, \ie
\begin{equation}
\bar{c} = \frac{\int_V c({\bf x})\,d^3x}{\int_V d^3x}= 
\frac{1}{V}\;  \int_V c({\bf x})\,d^3x
\end{equation}
The time update is accomplished through
a double integration of eq. (\ref{dcepiece2.eq}) over the cell volume
and over the time-step $\Delta t$. 
Upon performing such integrations,
we find
\begin{equation}
\bar{c}^{\tau+\Delta \tau}_i - \bar{c}^{\tau}_i = - \Delta \tau
\sum_{\ell=1,2,3} \left(\Phi_{x_\ell}^x-\Phi_{x_\ell-1}^x\right)
\end{equation}
where the index $\ell$ runs
over the three spatial coordinates and accounts 
for the flux through the computational cell boundaries 
along the three spatial directions.
The time-averaged flux in physical space is defined as
\begin{eqnarray} \label{fluxx.eq}
\Phi_{x_\ell}^x
& = & \frac{1}{\Delta t\Delta x} 
\int_t^{t+\Delta t}c_i\, u_\ell \;\vert_{x_\ell}\; dt^\prime
\end{eqnarray}
with the integrand evaluated
at the cell boundaries $x_\ell$. The numerical
solution $c(t,{\bf x},p)$ is reconstructed from the 
volume average values ($\bar{c}$) as a piece-wise linear 
interpolation. An important part of this step is the effort
to restrict the slopes of the linear interpolation. In fact,
having the potential to become artificially large near extrema or 
discontinuities, they can cause unwanted oscillations 
and render the scheme unstable. This problem is eliminated
by demanding that the interpolated function is monotone \cite{vanleer73}.
The reconstructed solution can then be upgraded
``exactly'' in Lagrangian coordinates as
$c(t^\prime,{\bf x},p)= c(t,{\bf x}-\int^{t^\prime}_t{\bf v}\,dt,p)$,
allowing the calculation
of the flux in eq. (\ref{fluxx.eq}) at each interface $x_\ell$.

\subsection{Injection} \label{inj.se}

The source terms in eq. (\ref{dce3.eq}) and (\ref{edce.eq})
are responsible for the variations
\begin{eqnarray}
\bar{n}^{\tau+\Delta \tau}_i - \bar{n}^{\tau}_i  & =  & Y_i  \\
\bar{\varepsilon}^{\tau+\Delta \tau}_i - \bar{\varepsilon}^{\tau}_i  & =  &
\Sigma_i
\end{eqnarray}
respectively, where
\begin{eqnarray}
Y_i = \frac{1}{V\Delta t}\int_t^{t+\Delta t}\int_V Q_i \;d^3x\,dt = 
 \frac{1}{\Delta t}\int_t^{t+\Delta t} \bar{Q_i} \;dt   \\
\Sigma_i = \frac{1}{V\Delta t}\int_t^{t+\Delta t}\int_V S_i \;d^3x\,dt = 
 \frac{1}{\Delta t}\int_t^{t+\Delta t} \bar{S_i} \;dt   \\
\end{eqnarray}
and $Q_i$ and  $S_i$ have been defined by eq. (\ref{source.eq}) and
(\ref{esource.eq}), respectively.
In this section, we shall focus our attention on source
contributions provided by CR injection 
at shocks. 
With this term we indicate the number of particles that upon passing through 
a shock are assumed to undergo the diffusive shock acceleration mechanism. 
Since the latter is treated as instantaneous, injection here basically refers
to the deposition in the post-shock region of CRs with a power-law distribution
in momentum space which accords with the diffusive shock acceleration theory
(see \S \ref{nuch.se}) for further details).
Next (\S \ref{prosec.se}), we will 
outline the scheme for the injection term due
secondary electrons produced in inelastic p-p collisions.

The scheme described here for the 
injection of CRs at shocks is
based on the {\it thermal leakage} model which 
seems to be observationally supported at least 
by {\it in situ} measurements
at the earth bow shock \cite{elmopa90}.
Here, we assume that upon passing through the shock
most of the gas thermalizes to a Maxwellian distribution 
with post-shock temperature $T_2$,
\ie 
\begin{equation} \label{maxw.eq}
f_{m}(p) = n_2\, (2m_p kT_2)^{-3/2}\,
\exp\left(-\frac{p^2}{2m_p kT_2}\right)
\end{equation}
with $n_2$ the total number of particles of the distribution. 
For a particle to return upstream
it is necessary not only 
that it propagates faster than the shock wave, but also that it 
has enough energy to escape ``trapping'' by Alfv\'en waves
generated in the downstream turbulence \cite{mavo95,mavo98}.
Thus, only those particles in the high energy tail of the thermal 
distribution 
will have a chance to re-cross the shock and get injected into
the acceleration mechanism.
The numerous, complicated 
details of the physics underlying 
the injection mechanism are conveniently modeled
by a few parameters \cite{joka93,beyeks94,beyeks96,kajo95}.
One of them, $c_1$,
defines the momentum threshold for the particles
of the thermal distribution to be injected, as  
\begin{equation} \label{pinj.eq}
p_{inj} = c_1\,2\sqrt{m_p kT_2}.
\end{equation}
In practice we assume that at $p_{inj}$  
the thermal distribution, $f_m(p_{inj})$ [eq. (\ref{maxw.eq})],
joins ``smoothly'' into 
the power-law distribution of the CRs. That implies 
\begin{equation} \label{powerlaw.eq}
f_{cr}(p) = f_{m}(p_{inj}) \;\left(\frac{p}{p_{inj}}\right)^{-q},
\end{equation}
where $q = 3r /(r-1)$, $r$ is the shock compression ratio 
and $p_{inj} \leq p\leq p_{max}$.
With these choices, $c_1$ is the only free parameter
in the injection model. In fact, 
the total number of injected particles
is given by
\begin{equation} \label{qinj.eq}
n_{inj} = \int_{p_{inj}}^\infty 4\pi\,p^2\,f_{cr}(p)\, dp = 
4\pi\,f_m(p_{inj})\,p_{inj}^3\,
\frac{\left(\frac{p_{max}}{p_{inj}}\right)^{3-q}-1}{3-q}.
\end{equation}
and the injection efficiency,
\ie the fraction of downstream thermal gas particles that 
are injected in the acceleration process, is 
[\cf eq. (\ref{maxw.eq}) and (\ref{pinj.eq})]
\begin{equation}  \label{etainj.eq}
\eta_{inj} \equiv \frac{n_{inj}}{ n_2}
= 8\,\sqrt{\frac{2}{\pi}} \, c_1^3\,e^{-2c_1^2}
\, \frac{\left(\frac{p_{max}}{p_{inj}}\right)^{3-q}-1}{3-q},
\end{equation}
practically independent of $T_2$.
In an alternative approach, where the downstream gas distribution
is not specified, $\eta_{inj}$ and $c_1$ can be taken as independent
parameters \cite{beyeks94,beel99}. With the above setting,
CR particles are injected per unit shock surface at a rate 
\begin{equation}
\bar{Q}_i = \eta_{inj}\; \frac{\rho_1\,u_s}{m_p} =  
\eta_{inj}\; \frac{n_2\,u_2}{r}
\end{equation}
where $n_2\,u_2=n_1\,u_1$ is the number flux of thermal 
particles impinging on the shock.
Given the injection rate, $Q_i$, and the distribution 
of the injected particles in eq. (\ref{powerlaw.eq}),
the energy injection term, $S_i$, can be easily
computed as well by means of eq. (\ref{esource.eq}).

The actual values of $c_1$ and $\eta_{inj}$ are not known
and most likely they are not constant, but a function of the conditions
of the flow.
For ionic CRs, observations at the earth's bow
shock indicate that an injection efficiency, $\eta_{inj}$,
can be as high as
$1.25 \times 10^{-3}$ \cite{lee82} or even 
$\sim 10^{-2}$ \cite{elmopa90}. In theoretical 
studies of shock acceleration at
supernova remnants the parameter $c_1$ is assumed in the range 2.3-2.5
corresponding to values of $\eta_{inj}$ ranging between 
a few $\times 10^{-3}$ to $10^{-4}$
\cite{drmavo89,joka93,kajo95,beyeks94,mavo95,beyeks96,beel99}.
It is important to notice that
the value of $c_1$ (in addition to $p_{max}$ and the slope $q$),
regulates the amount of flow kinetic energy
that is transferred to the CRs.
CR protons can be dynamically important by 
exerting a pressure 
\begin{equation} \label{pcr1.eq}
P_{cr} = \frac{4\pi}{3}\; \int_{p_{inj}}^{p_{max}}
f(p)\,p^3\,v\;dp
=\frac{4\pi}{3}\; c\,\int_{p_{inj}}^{p_{max}}
f(p)\,\frac{p^4}{(m_p^2 c^2+p^2)^{1/2}}
\;dp.
\end{equation}
The ratio of the CR pressure
to the ram pressure of the flow, $\rho_1 \,u_1^2$, 
can be regarded as a first order indication of
the relative importance of the two components.
For a flat CR distribution, \ie $q\simeq 4$, which is
typical of the cosmological case \cite{mint00,mrkj01},
by using the above expression for $P_{cr}$ and 
neglecting the sub-relativistic contribution we find 
\begin{equation}\label{pcram.eq}
\frac{P_{cr}}{\rho_1 u_1^2} =
\frac{8}{3}\sqrt{\frac{2}{\pi}} \, c_1^3\,e^{-2c_1^2}
\,\left(\frac{m_pc}{p_{inj}}\right)^{3-q}
\, \left(\frac{c}{u_s}\right)^2 \,
\frac{\left(\frac{p_{max}}{p_{inj}}\right)^{4-q}-1}{4-q}
\end{equation}
Thus, 
for a CR ion component extending between 1 GeV and $10^6$ GeV,
appropriate for shocks in galaxy clusters,
the back-reaction of the accelerated particles would not be negligible
with the aforementioned choices of the parameter $c_1$.
It appears that in cosmological simulations of large scale structure 
formation, most of the kinetic energy is processed by shocks with Mach number
in the range 3-6 \cite{minetal00}. According to 
recent results from a newly developed 
Adaptive Mesh Refinement scheme with a 
resolution down to the diffusion 
length scale and capable of 
including self-consistently the dynamical role 
of the accelerated particles \cite{kjlvs00}
shocks with Mach number $M\sim 3-6$ are expected to be 
``moderately'' efficient (up to 30\%) but not strongly modified.
Thus, the effect of the back-reaction 
can be emulated by slightly increasing the value of $c_1$,
which has the effect of reducing the total number of injected CRs.
For the above range of flow parameters, we find that a 
value of $c_1=2.6$ produces an injection efficiency $\eta_{inj}$ 
consistent with above nonlinear calculations \cite{kjlvs00,beksye95}.
In general, however, one must be very cautious about the 
choice of $c_1$. Its value should be determined on a 
individual basis, based on the properties
of the simulated shocks where most of the CRs are being produced
and in accord with the possible non-linear behavior of the
shock acceleration mechanism there. 

The physical process of injection of CR electrons is
more complicated and basically not yet fully understood.
Part of the reason is due to the fact that electrons
carry only a small fraction of the momentum of the flow
and, therefore, they do not affect the dynamics of the shock.
Thus, unlike the ions, whose behavior can be constrained 
by general considerations of energy and momentum conservation,
the electrons behave as test particles and their dynamics
is determined by the details of the plasma wave-particle interactions.
The latter is very difficult to model appropriately in 
computer simulations. In addition, until recently 
the observational results available in this respect were 
very limited. In this situation, the progress in the
understanding of CR electron injection at shocks has been 
relatively slow. However, for the present purpose of a
numerical treatment of CR injection at cosmic shocks,
a viable and reasonable approach is to assume that 
the ratio of the cosmic
ray electrons to ions at relativistic energies is 
fixed to a value $R_{e/p}$ \cite{elbeba00}. 
This phenomenological approach is supported by 
some observational evidence from experiments with Galactic
cosmic rays indicating that this ratio is possibly 
in the range 1-5 \% \cite{muta87,mulletal95}.

\subsection{Production of Secondaries} \label{prosec.se}

Electrons and ions injected at shocks
provide the main source of CRs and 
are usually referred to as {\it primary}. 
As high energy protons propagate through the galactic or
intergalactic medium, they collide with the background thermal protons
and, from the hadronic interaction, secondary products are generated.
The latter include photons, leptons and hadrons and, therefore,
might be an important source of relativistic electrons.
The main channels for the production of secondary electrons (positrons) 
are \cite{gaisser90}:
\begin{eqnarray} 
p + p & \rightarrow & \pi^\pm + X  \label{pppi.eq}\\
p + p & \rightarrow & K^\pm + X. \label{ppk.eq}
\end{eqnarray}
\noindent
where $X$ indicates all the by products of the reactions.
Charged pions decay with a 
lifetime of $2.6 \times 10^{-8}$ s \cite{pdg00}, mostly into
\begin{eqnarray} 
\pi^+ & \rightarrow & \mu^+ + \nu_\mu  \label{pi+d.eq}\\
\pi^- &\rightarrow &\mu^- + \bar{\nu}_\mu  \label{pi-d.eq}
\end{eqnarray}
\noindent
Kaons, analogously, have a lifetime if $1.24 \times 10^{-8}$ s and
decay primarily into muons (63.5\%) and pions (21.2\%) \cite{pdg00}:
\begin{eqnarray} 
K^+ & \rightarrow & \mu^+ + \nu_\mu \label{kd1.eq}\\
K^- & \rightarrow & \mu^- + \bar{\nu}_\mu \label{kd12.eq}\\
K^\pm & \rightarrow & \pi^0 + \pi^\pm   \label{kd2.eq}
\end{eqnarray}
\noindent
The relative contributions of the two channels
($p+p \rightarrow \pi^\pm +X $ 
and $p+p \rightarrow K^\pm +X$) are energy dependent. 
So, for example, the fraction of secondary muons from $K$ decay is 
8\% at about 100 GeV, 19\% at 1 TeV and approaches 
asymptotically 27\% at higher energy \cite{gaisser90}.
In turn, muons have a lifetime of $2.2 \times 10^{-6}$ s \cite{pdg00} before 
decaying into
\begin{eqnarray}  
\mu^+ & \rightarrow & e^+ + \nu_e + \bar{\nu}_\mu \label{mu+d.eq} \\
\mu^- &\rightarrow& e^- + \bar{\nu}_e + \nu_\mu . \label{mu-d.eq}
\end{eqnarray}
\noindent
In addition to $p+p$ inelastic collisions, 
the above cascades are also triggered by the interaction of
$p+$He, $\alpha+$H and $\alpha+$He, which, for example, increase
the overall yield of secondary $e^\pm$
by a factor 1.4 for a metal composition relative to the interstellar 
medium \cite{dermer86}. 

In general we can write the production spectrum of
secondary electrons as \cite{most98}
\begin{equation}  \label{emissnt.eq}
j_s(\varepsilon_s) = n_H \: \sum_{i = \pi, K} 
\int_{\varepsilon_p^{min}}^\infty
d\varepsilon_p \; J_p(\varepsilon_p)\, 
\langle \zeta\sigma_i(\varepsilon_p) \rangle \:
\int_{\varepsilon_i^{min}(\varepsilon_s)}^{\varepsilon_i^{max}(\varepsilon_p)}
d\varepsilon_i \: F_s(\varepsilon_s, \varepsilon_i) 
F_i(\varepsilon_i, \varepsilon_p)
\end{equation}
\noindent
where $J_p(\varepsilon_p) $ is the proton flux; 
$\langle \zeta\sigma_i(\varepsilon_p) \rangle$ is the inclusive cross
section\footnote{Inclusive means that which describes the process
\begin{equation} 
p+p \rightarrow i + X
\end{equation}
where $i$ is in general a secondary particle}
of the processes (\ref{pppi.eq}) and (\ref{ppk.eq});
$\varepsilon_p^{min}$ is the minimum proton energy required to 
produce a meson of energy $\varepsilon_i^{min}$ and $\varepsilon_i^{max}$
the maximal energy of the produced meson;
$\varepsilon_i^{min}$ in turn is the minimum required energy of a meson 
for production of secondaries of energy $\varepsilon_s$; finally 
$F_i(\varepsilon_{\pi,K}, \varepsilon_p)$ 
are the spectra of $\pi$ and $K$ produced from the collision of a proton of
energy $\varepsilon_p$
and $F_s(\varepsilon_s, \varepsilon_{\pi,K})$ the distribution of secondaries 
from the subsequent decay of the above collision products.
For a full summary of the technique employed here in calculating the cross
sections and the functions $F_s$ and $F_i$ we refer to ref. \cite{most98}.

\subsection{Energy Losses} \label{radlos.se}

We have considered the energy losses due to a variety of 
physical mechanisms that become relevant at different energy regimes.
For the electrons, the most effective process is due to 
Coulomb losses in the low energy end and synchrotron and inverse Compton 
emission at high energies. Bremsstrahlung losses are also included
for completeness, although less relevant.
For ion CRs, Coulomb losses, which are mechanical 
in nature, are dominant 
below relativistic energy. However, given the much smaller 
cross section for radiative losses for the ions ($\propto 1/m_p^2$)
as compared to electrons ($\propto 1/m_e^2$), the next important 
loss mechanism for the CR ions 
beside adiabatic expansion is that due to inelastic collisions
with the thermal background nuclei. 
In the following we shall provide the functions
\begin{equation}
 b(p) \equiv \frac{dp}{dt}
\end{equation}
and the associated cooling time 
\begin{equation}
\tau_{cool} \equiv \frac{p}{b(p)}
\end{equation}
relative to each relevant process, and
separately for electrons and ions.

\subsubsection{Electrons}

Losses due to Coulomb collisions are expressed by \cite{stmo98}
\begin{eqnarray}  \label{coul.eq} \nonumber
\left(\frac{dp}{dt}\right)_{Coul} & = &
\frac{2\pi Z^2e^4}{m_ec^2}\,n\, \left\{
\ln\left(\frac{m_e^3c^4}{4\pi e^2\hbar^2Z}\right)+
\ln\left[\frac{(1+\phat^2)^{1/2}}{n}\right]-\frac{3}{4}\right\}  \\
  & = &
3.01\times 10^{-29} \left\{1+[\ln(1+\phat^2)^{1/2}-\ln n]\frac{1}{73.56}
\right\}\;n
\;\mbox{erg\,cm}^{-1}
\end{eqnarray}
where $Z$ is the electric charge of the ion species,
$n$ is the number density of the background
gas in cm$^{-3}$ and $\hat{p}\equiv p/m_ec$.
Taking $Z=1$, the corresponding cooling time is defined as
\begin{equation}  \label{coult.eq}
\tau_{Coul}   = 2.53 \times 10^9 \,
\left(\frac{\phat}{100}\right) \,
\left(\frac{n}{10^{-3}{\rm cm}^{-3}}\right)^{-1} ~~~ {\rm yr}.
\end{equation}
Bremsstrahlung losses are defined as 
\cite{ginzburg79,stmo98}
\begin{eqnarray} \label{bremm.eq} \nonumber
\left(\frac{dp}{dt}\right)_{brem} & = &
4\alpha_f\, r_e^2\,\gamma
m_ec^2\,Z(Z+1)\,n\left[\ln(2\gamma)-\frac{1}{3}\right]
\\
& = &
3.8 \times 10^{-33}\left\{\ln[2\,(1+\phat ^2)^{1/2}
]\,-\frac{1}{3}\right\}\, \phat\, n
\;\mbox{erg\,cm}^{-1}
\end{eqnarray}
where $\alpha_f$ is the fine structure constant, $r_e$ the classical 
electron radius. Such mechanism is basically unimportant
as it becomes effective on a time-scale
\begin{equation}  \label{bremt.eq}
\tau_{brem} = 6 \times 10^{12} \,
\left(\frac{n}{10^{-3}{\rm cm}^{-3}}\right)^{-1} ~~~ {\rm yr},
\end{equation}
again taking $Z=1$.
The other relevant loss mechanisms for electrons are due to 
synchrotron emission and inverse Compton scattering 
off the cosmic microwave background 
photons. After averaging over the electrons pitch angle,
their combined contribution is given by
\begin{eqnarray} \label{syncic.eq} \nonumber
\left(\frac{dp}{dt}\right)_{sync+IC} & = &
\frac{4}{3}\frac{\sigma_T\,p^2}{m_e^2c^2}\, (u_B+u_{cmb}) \\
  & = &
 8.94\times 10^{-25} (u_B+u_{cmb})\, \phat^2 \;\mbox{erg\,cm}^{-1}
\end{eqnarray}
where $U_B$ and $u_{cmb} = 4.2\times 10^{-13}\,(1+z)^4$ 
are the energy density in magnetic field and 
cosmic microwave background at a given cosmological red-shift, $z$, 
respectively, both in units of erg\,cm$^{-3}$.
For high energy electrons this is the most severe energy loss
mechanism with a typical time-scale at red-shift $z=0$
\begin{equation}  \label{ict.eq}
\tau_{sync+IC} 
= 2.3 \times 10^{8} \,
\left(\frac{\phat}{10^4}\right)^{-1} \,
\left(1+\frac{u_B}{u_{cmb}}\right)^{-1} ~~~ {\rm yr}.
\end{equation}

\subsubsection{Ions}

Coulomb collisions are efficient at a rate \cite{stmo98}
\begin{eqnarray}  \label{coulp.eq} \nonumber
\left(\frac{dp}{dt}\right)_{Coul} & = &
\frac{2\pi Z^2e^4}{m_ec^2}\,n\, 
\ln\left(\frac{\gamma^2m_e^3c^4}{\pi e^2\hbar^2n}\;
\frac{m_p\beta^4}{m_p+2\gamma m_e} \right)  \frac{\beta^3}{x_m^3+\beta^3} \\
  & = &
3.01\times 10^{-29} \left\{1+\left[
\ln\left( \frac{\phat^4/(1+\phat^2)}{1+2(m_e/m_p)(1+\phat^2)^{1/2}}\right)
-\ln n\right]\frac{1}{75.7}
\right\} \cdot\nonumber  \\&   &
\frac{\beta^3}{x_m^3+\beta^3}\;n
\;\mbox{erg\,cm}^{-1}
\end{eqnarray}
where $beta = v/c$,
\begin{equation}
x_m=\left(\frac{3\pi}{4}\right)^{1/3}\, \left(\frac{2kT_e}{m_ec^2}\right)^{1/2}
=6.4 \times 10^{-2}\, \left(\frac{T_e}{10^7\mbox{K}}\right)^{1/2}
\end{equation}
and we now define $\hat{p}\equiv p/m_p c$.
For protons in the low energy end within 
the range considered here, Coulomb collisions
are the only significant energy loss mechanism,  
with a time-scale comparable to the Hubble time, 
\begin{equation}  \label{brempt.eq}
\tau_{Coul}   = 4.8 \times 10^{9} \,
\left(\frac{\phat}{0.1}\right) \,
\left(\frac{n}{10^{-3}{\rm cm}^{-3}}\right)^{-1} ~~~ {\rm yr},
\end{equation}
where $Z=1$ is used.
Photo-pair and photo-hadron production of CR ions $A$,
interacting with the cosmic microwave background photons, \ie
\begin{eqnarray}
A+\gamma_{cmb} &\rightarrow & A+e^++e^- \\
A+\gamma_{cmb} &\rightarrow & A+\pi 
\end{eqnarray}
are mostly negligible at the energies we include. 
In fact, for red-shift $z= 0$ and taking the average CMB 
photon energy $\langle \epsilon_{cmb} \rangle = 7\times 10^{-4}$eV,
the thresholds for the above reactions,
in terms of the Lorentz $\gamma$ factor, are
respectively
\begin{eqnarray}
\gamma_{min} = \frac{m_ec^2}{\langle \epsilon_{cmb} \rangle} \simeq 7\times 10^8\\
\gamma_{min} = \frac{\xi m_\pi c^2}{2\langle \epsilon_{cmb} \rangle}\,
\left((1+\frac{\xi m_\pi}{2m_p}\right) \simeq \xi \times 10^{11}
\end{eqnarray}
reachable only for ultra high energy CRs
(where $\xi$ is the multiplicity of the produced pions).
Rather, inelastic 
collisions of CR ions off nuclei of the thermal 
background gas are more significant, being at a rate
\cite{masc94}
\begin{eqnarray}  \label{p-p.eq} \nonumber
\left(\frac{dp}{dt}\right)_{p-p} & = & \sigma_{\pi,inel}
(E_p-m_pc^2) n \\
 & = & \left\{ \begin{array}{lll} 2.91\times 10^{-29}\,(\phat - \beta)n
\;\mbox{erg\,cm}^{-1}  & \mbox{if} & e_p\geq 1.22 \mbox{GeV} \\
  0 &  \mbox{if} & e_p < 1.22 \mbox{GeV} 

\end{array} \right.
\end{eqnarray}
and characterized by a energy loss time-scale
\begin{equation}  \label{p-pt.eq}
\tau_{p-p}  = 5.5 \times 10^{10} \,
\left(\frac{n}{10^{-3}{\rm cm}^{-3}}\right)^{-1} ~~~ {\rm yr}
\end{equation}

\section{Test Results of the Code} \label{tere.se}

In the following we present a set of
numerical experiments that test the performance of the code
in various cases of direct interest. 
In particular we consider the 
time evolution of an initial distribution of 
CR protons and electrons as they lose energy 
subject to mechanisms described in \S \ref{radlos.se}.
Alternatively, we also present
the evolution toward the steady state of 
electrons which are continually injected as secondary products 
of p-p collisions and at the same time lose energy due to 
Coulomb and inverse Compton losses.
For the first experiment we take
the initial spectrum of the cosmic
rays (protons and electrons) 
as a power-law with logarithmic slope $q_0 = 4.3$.
A similar distribution (but with a different normalization)
will also provide the parent CR protons that produce the
secondary electrons in the last example.
The numerical solution, which gives the number density
of CRs in each momentum bin as a function of time, $t$,
is plotted and compared with the analytical solution. 
The latter is obtained by integrating between 
each momentum bin bounds the exact, analytical 
CR distribution corresponding to the time $t$ \cite{gisy64},
for the same initial distribution evolved by the code.
As for the Coulomb, bremsstrahlung and p-p collision losses,
we assume
that the CR protons or electrons propagate through a medium
with number density $n_{gas} = 10^{-3}{\rm cm}^{-3}$, 
typical of the core of clusters of galaxies. In order to further
mimic cosmic conditions, the test are run for a time 
$\tau_H = 1.5 \times 10^{10}$ yr so that all the cooling
time-scales of relevance in a cosmological simulation are 
included.

\subsection{Evolution of Cosmic Ray Power-Law Distributions} \label{evpld.se}
\begin{figure}
\vskip 9.5truecm
\plotfiddle{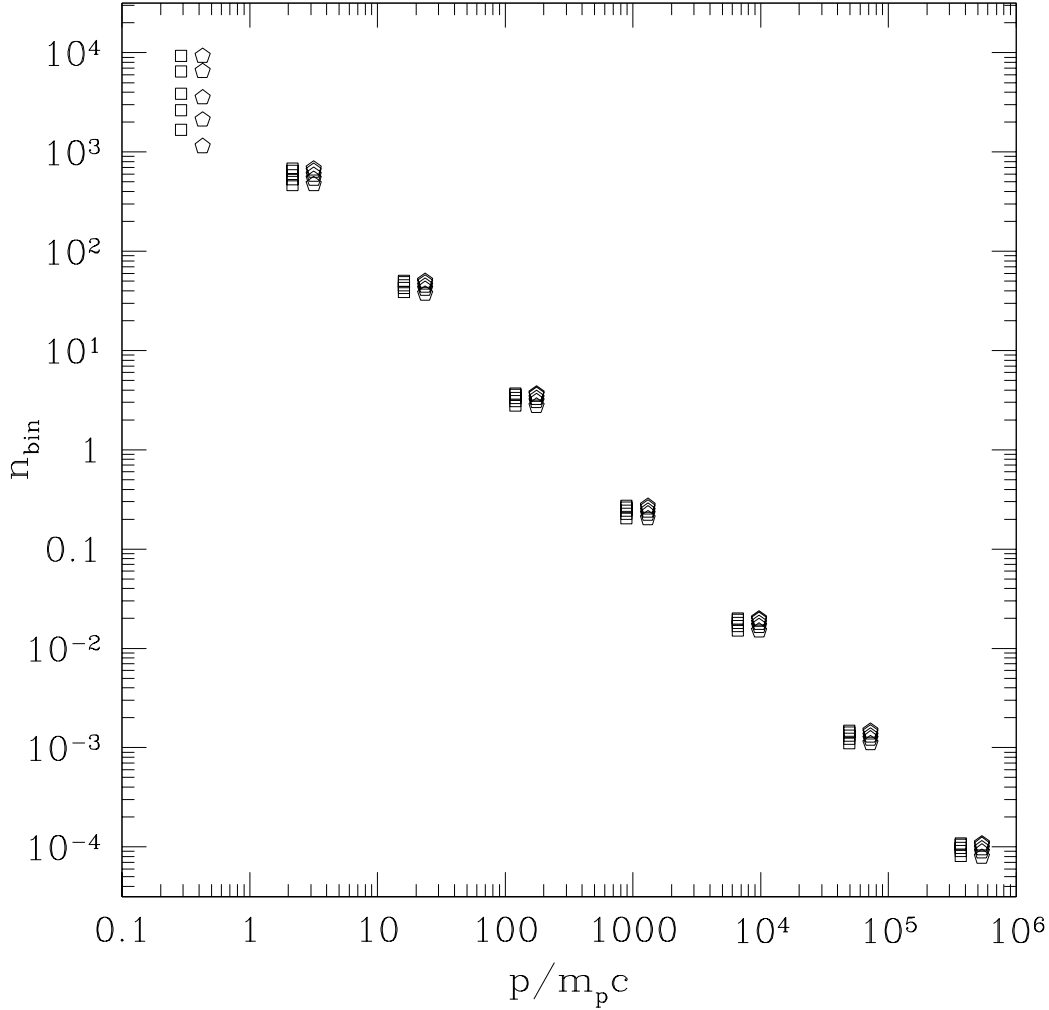}{110pt}{0}{70}{70}{-220}{-90}
\caption{Evolution of CR proton spectrum with initial logarithmic slope
$q_0 = 4.3$, subject to Coulomb losses and 
p-p inelastic collisions over a period of 15 Gyr.}
\label{pcrev.fig}
\end{figure}
In fig. \ref{pcrev.fig} we present the temporal evolution of a 
distribution of CR protons subject to Coulomb losses and
p-p inelastic collisions. 
For this experiment we use 8 momentum bins, with the 
CR spectrum extending between $p_0= 0.1$GeV$/c$ up to 
$p_8=10^6$GeV$/c$ ($c$ is the speed of light). 
The distance between subsequent bin bounds is 
constant on a logarithmic scale and is $\Delta w =[\log(p_8)-\log(p_0)]/8 = 
0.875$. 
Since the losses 
are only mild in this case,
the numerical evolution of the CRs is computed by the 
scheme presented in \S \ref{admo.se}.
For each momentum bin, on
the ordinate axis we plot the number density
of CR protons as computed from the code (pentagons)
and from the analytical solution (square). 
Here and in the following figures,
in order to facilitate the visual comparison of the results,
for each bin the squares are drawn in correspondence of 
the middle of the bin and pentagons at a slightly 
higher momentum value, although the bins are identical 
for both cases. 
The points in the plot correspond to the following 
times $t/\tau_H = 0.1,0.3,0.5,0.8,1 $.
The time sequence is such that, for each bin,
points in the upper positions
correspond to earlier times.
\begin{figure}
\vskip 9.5truecm
\plotfiddle{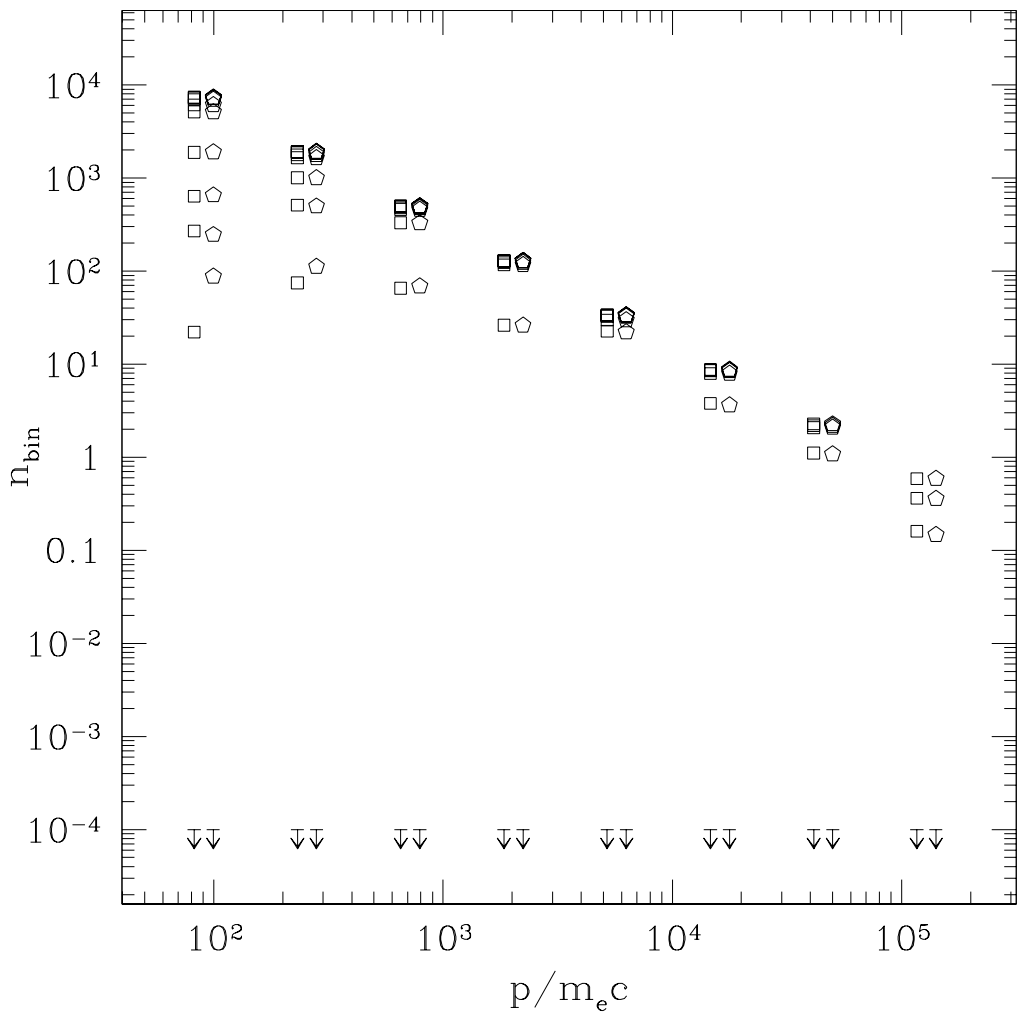}{120pt}{0}{70}{70}{-220}{-90}
\caption{Evolution of CR electron spectrum subject to Coulomb,
bremsstrahlung and inverse Compton losses over a period of 15 Gyr.
The initial logarithmic slope was set to 4.3}
\label{ecrev.fig}
\end{figure}
As we can see, above 1 GeV the CRs are only sensitive to 
the p-p losses which deplete all the momentum bins at 
the same rate. In this regime the two solutions,
the numerical and the analytical one, are, for 
all practical purposes, identical.
At lower energies, \ie for the first bin, 
the numerical solution still 
follows very accurately the exact solution. Only in the last
two points of the time sequence, 
after $t = 1.2\times 10^{10}{\rm yr}= 0.8\;  \tau_{Coul} \; (\phat/0.3)$, 
the numerical solution cools a little bit faster than the 
analytical one and the number of CR protons in the first bin (only!)
is slightly underestimated by a factor $\lesssim 1.4$.

Analogously, in Fig. \ref{ecrev.fig}, we present a comparison 
between the numerical (pentagons) and the analytical (square) 
solutions for
the evolution of an initial power-law distribution of CR electrons.
In this case, the numerical solution of the
evolution of the CR distribution is computed
by means of the more sophisticated (and more expensive)
scheme outlined in \S \ref{admoe.se}.
Again we use 8 momentum bins, but the extrema of the
CR electrons spectrum are now $p_0= 50 $MeV$/c$ and
$p_8=2\times 10^5$MeV$/c$ with a distance between 
subsequent bin bounds $\Delta w = 0.45$. 
The inverse Compton losses are computed by setting 
$u_b \ll u_{cmb} =  4.2\times 10^{-13}$.
The values of the CR number density for each bin
in Fig. \ref{ecrev.fig} has been plotted for the following 
times (ages) of the distributions: $t/\tau_H =
5\times 10^{-4},1\times 10^{-3},3\times 10^{-3},1\times 10^{-2},
2\times 10^{-2},0.1,0.25,0.45,0.6,.7 $.
The above times have been chosen in order to test the 
accuracy of the numerical routine in different regimes.
For example, the first three times correspond to a 
fraction of $\tau_{IC+sync}$
for particles with $\phat\simeq 10^4$, whereas the last times
are a significant fraction of $\tau_{Coul}$. 
\begin{figure}
\vskip 9.5truecm
\plotfiddle{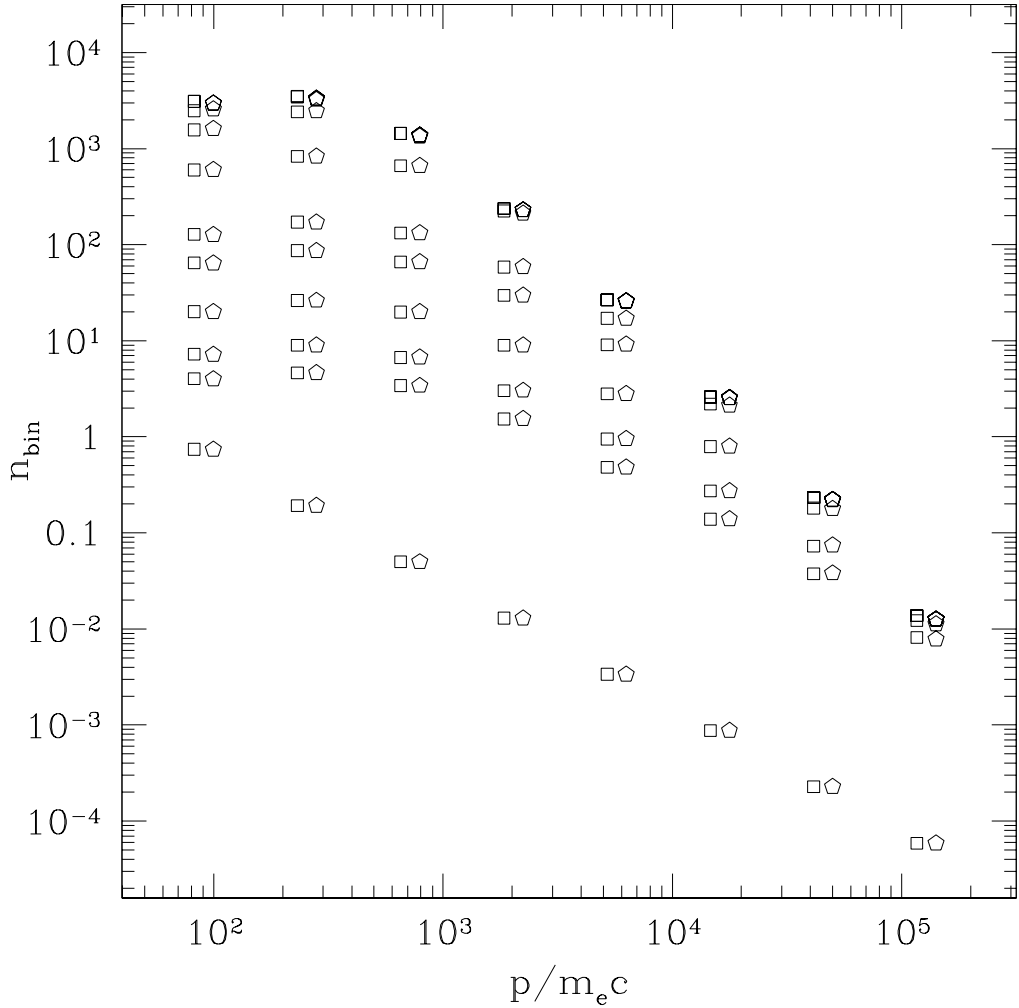}{120pt}{0}{70}{70}{-220}{-90}
\caption{Evolution of secondary CRs produced in p-p collision
by a power-law distribution of parent CR protons with 
logarithmic slope was set to 4.3}
\label{screv.fig}
\end{figure}
Again, since the evolution of the CR spectrum is 
determined by energy losses, 
the time evolution is such that the upper points
correspond to earlier stages. 
Since the cooling time for the CR electrons in the high energy bins 
is much shorter than the total evolution time $\tau_H$, eventually
such bins will be evacuated. Since the plotted dynamical
range is limited, we adopt ``upper limit'' symbols for those
bins where cooling has reduced the number of particles below
an arbitrary numeric level of $10^{-4}$.
We notice once again a very accurate correspondence between the 
code and the analytical solution. As for the CR protons, 
the numerical solution
departs slightly from the analytical one at the lower energies (for the 
first 2 bins). However, this now takes place 
only after $t \ge  0.6 \, \tau_H = 7.1\, \tau_{Coul}$,
to be compared with $t\ge 0.8 \tau_{Coul}$ for the protons case.
This means that the present 
scheme is more accurate not only for the evolution of
the high energy part of the spectrum but also the low energy end.
This feature is relevant in the context of cosmological simulations
because the time-scales for the energy losses of electrons at all energies
of interest are significantly shorter than a Hubble time.
Notice that, unlike the previous case, the numerical solution
now overestimates the total number of CRs in the lower momentum 
bins. The discrepancy, however, appears only when a momentum
bin has already been significantly evacuated of electrons and is
close to complete depletion. From 
Fig. \ref{ecrev.fig} we can see that the difference between the 
true and the numerical solution (for both first and second bins) 
is only a small fraction of the difference between the initial 
and the current value. Also, notice that
after the second momentum bin has been evacuated, \ie 
around $t/\tau_H \sim 0.5$,
the scheme is working with only one momentum bin. 

Finally, in Fig. \ref{screv.fig} we compare the numerical 
and analytical results for the evolution of a distribution of 
secondary CR electrons, continuously injected through the processes
described in \S \ref{prosec.se}. 
In this case we start from a low
background distribution of CR electrons
and the continuous injection of new particles builds up the steady state
distribution. Thus, unlike in the previous examples, now the
time evolution is such that the lower points correspond to earlier 
times. The CR electron distributions have been plotted for
both  the numerical (pentagons) and analytical (square) solutions 
for the following times: $t/ \tau_H=
5\times 10^{-4},1\times 10^{-3},3\times 10^{-3},1\times 10^{-2},
2\times 10^{-2},0.1,0.3,0.5,0.8,1$. Again the numerical 
scheme follows very precisely the evolution
of the spectrum of the secondary electrons.
Since the slope of the parent CR protons was $q_0=4.3$, 
the injection spectrum has a logarithmic slope $q_i=4.3$ \cite{masc94}.
Thus, as expected, the slope of the steady state spectrum is respectively
steeper at high energies and flatter at the low energy end 
than the injection spectrum by one unity.
We notice, however, that the first and last bins are even 
flatter and steeper respectively than the above prediction.
At low energies, this is a physical effect due to the reduction
of the injection rates of secondary electrons as
we approach the lower energy limit for electron 
production $\sim m_\mu/2\sim 50$MeV \cite{bardon65}.
The steepening of the slope of the last bin, on the other hand, is a 
boundary effect, namely the lack of incoming flux
of electrons from higher energies.
This is due to the fact that we included an injection 
spectrum only up to electron energy of 100 GeV.
Since the same injection spectrum was used for both the numerical
and the analytical solutions, the test comparison is unaffected 
by this choice. 

\subsection{Cosmic Ray Distributions in a Galaxy Cluster
from a Cosmological Simulation}

\begin{figure}
\vskip 5.5truecm
\caption{Gray-scale image of a 2-D slice of the 
gas density distribution in units cm$^{-3}$ for a cluster selected 
from the simulation. Linear size of the image is ca. 5 \hinv Mpc.}
\label{rho.fig}
\end{figure}
At last we present the distribution of CR protons as well as primary 
and secondary electrons, produced in a cosmological simulation of 
large scale structure formation performed by implementing
COSMOCR into a hydro+N body
cosmological code \cite{rokc93}.
The simulations are described in
detail elsewhere \cite{mint00,mjkr01,mrkj01}. 
For the present purpose it suffices to mention that
we assumed a SCDM cosmology with
an initial power spectrum 
of density fluctuations with cluster normalization
$\sigma_8 = 0.6$ and 
spectral index $n$ = 1; normalized Hubble constant 
$h \equiv$ H$_0$/(100 km s$^{-1}$ Mpc$^{-1}$) = 0.5; 
total mass density $\Omega_M = 1$; and baryonic fraction $\Omega_b = 0.13$. 
For the simulation we selected a cosmological volume defined by
a cubic region of comoving size 50 \hinv Mpc and use 
$256^3$ cells for the baryonic matter and $128^3$ dark matter particles.
This corresponds to a spatial resolution of $\sim$200\hinv kpc.
The CR protons and primary electrons have been injected
and accelerated at shocks formed during the simulation
with the prescription described in \S \ref{crco.se},
whereas secondary electrons are produced in p-p collisions.

We have selected one of the clusters formed in the simulations.
In Fig. \ref{rho.fig} we show a gray-scale image of 
a slice of the gas density in units cm$^{-3}$ 
through the core of such clusters.
For the same cluster, in Fig. \ref{figp.fig} (left panel) 
we present the analogous distribution of CR protons, 
and in Fig. \ref{figes.fig} (left panels) 
that of primary (top) and secondary electrons (bottom) as well.
In addition to the CR spatial distribution for each species 
we also project onto the corresponding right panel 
the CR {\it spectral} distribution 
(upper-half) and log-slope (lower-half) for two locations
in the intra-cluster medium: the cluster center (open squares)
and its periphery nearby an accretion shock (filled pentagons).

\begin{figure}
\vskip 5.5truecm
\caption{Left: distribution of the total number density 
of CR protons in units of cm$^{-3}$.  
The length of a side of the image is 5 \hinv Mpc.
Right: number density (top) and power law index (bottom)
for each momentum bin for two locations in the intra-cluster 
medium, at the cluster core (open squares) and periphery 
(filled pentagons).}
\label{figp.fig}
\end{figure}
\begin{figure}
\vskip 11.5truecm
\caption{Same as in Fig. \ref{figp.fig} but for primary
(top panels) and secondary electrons (bottom panels).}
\label{figes.fig}
\end{figure}
The spatial distribution of each species resembles that
of the gas density. In addition, CRs are only present 
within the shocked regions. This makes sense because 
roughly speaking according to our injection mechanism 
the CR particles are generated downstream of shocks and are then
passively advected with the flow.
For the case of the protons (Fig. \ref{figp.fig}) 
the number density is higher in the cluster core 
(right panel - open squares) than 
in the periphery (right panel - filled pentagons)
due to the adiabatic compression undergone by the 
CRs together with the gas during the formation of the cluster.
And in fact, in the inner regions of the cluster where the 
gas is denser and, therefore, the Coulomb losses stronger,
the CR proton spectrum shows a mild flattening at low energies
(Fig. \ref{figp.fig}, right-panel bottom-section), with respect 
to the distribution in the outer region.

Analogously, the {\it total} number of 
CR primary electrons is larger 
at the cluster core than at its periphery
(Fig. \ref{figes.fig}, left-top panel). However, 
at the high energy end the primary electrons 
in the cluster center have been completely depleted due
to severe inverse Compton losses, unlike in the peripheral region
behind the accretion shock where fresh CR electrons
are being generated. Such difference is reflected 
also in the values of the log-slope of the distributions, 
steep for $\phat \ge 10^3$ in the cluster center and 
still relatively flat at the periphery near the shock
(Fig. \ref{figes.fig}, left-top panel).
Finally, the bottom part of Fig. \ref{figes.fig}
shows the properties of CR secondary electrons. 
Again the number density is higher at the center of the
cluster than at its outskirts. Notice that the number density
of secondary electrons roughly scales as $n_{p}n_{gas}\propto
n_{p}^2$. In fact, this qualitative difference is 
reflected, approximately in the right proportion,
in the larger
distance between the two density curves in the bottom right panel 
(upper-half) of Fig. \ref{figes.fig}, 
as compared to those in Fig. \ref{figp.fig}.
Notice also that the log-slope of the distribution 
of the secondary spectrum is steeper (flatter) 
than that of the parent CR protons at high (low) 
energies by approximately one unit 
as expected. Again, the first and last bin
make an exception to this expectation, in accord to our
explanation in \S \ref{evpld.se}.

\section{Conclusions}

In this paper we described COSMOCR, a 
code for cosmic ray associated
studies in cosmological simulations. 
We have described the scientific 
motivations behind such code, the challenges posed by the 
realization of a suitable algorithm and the various 
numerical schemes adopted here in order to follow the evolution
of different CR species. We have shown in \S \ref{tere.se} 
that the present code is able to follow with high accuracy
the evolution of a CR distribution
of protons as well as primary and secondary 
electrons subject to the mechanisms of energy 
loss of interest in the intra-cluster medium.
Additionally, COSMOCR has been implemented 
in a cosmological code \cite{rokc93}. The 
resulting distribution of CR populations are in accord
with what we know about the properties of the cosmic shocks 
responsible for the acceleration of such CRs \cite{minetal00}, 
and with our knowledge of the transport and losses 
mechanism that regulate their evolution.

\ack

I am indebted to T. W. Jones, H. Kang and D. Ryu for 
valuable comments on the manuscript. I am also grateful to an anonymous referee 
for providing useful comments. 
I wish to acknowledge support from 
a Doctoral Dissertation Fellowship 
at the University of Minnesota and from a fellowship
provided by the Research Training Network of the 
European Commission for the Physics of the Inter-galactic Medium.
This work has been supported in part by NASA grant NAG5-5055,
by NSF grants AST96-16964 and AST00-71167, by the University of Minnesota
Supercomputing Institute and by the Max-Planck-Institut f\"ur Astrophysik.

\newpage
\bibliographystyle{prsty}
\bibliography{papers,books,proceed}

\end{document}